# Lagrangian analysis of the laminar flat plate boundary layer

Mohammad Gabr*

*EgyptAir Maintenance & Engineering, Cairo, Egypt*

**The flow properties at the leading edge of a flat plate represent a singularity to the Blasius laminar boundary layer equations; by applying the Lagrangian approach the leading edge velocity profiles of the laminar boundary layer over a flat plate are studied. Experimental observations as well as the theoretical analysis show an exact Gaussian distribution curve as the original starting profile of the laminar flow. Comparisons between the Blasius solution and the Gaussian curve solution are carried out providing a new insight into the physics of the laminar flow.**

## Nomenclature

$a$ = total acceleration integrated over a distance x

$a_{inst.}$ = instantaneous acceleration

$\delta$ = displacement Thickness of the Boundary Layer

$\delta_{99}$ = thickness of the boundary layer at $\frac{u}{U_\infty} = 0.99$

$\delta_T$ = total displacement

$D$ = the drag force

$t$ = time

$u$ = velocity at certain y

$U$ = flat plate velocity

$U_\infty$ = free Stream velocity

$x$ = distance from the leading edge

$y$ = distance normal to the flat plate

$\Delta t$ = time interval between Time Lines

$\eta$ = non dimensional parameter

$\rho$ = density

*Email address for correspondence: Mohamed.Gabr@Polymtl.ca

$\mu$     =   dynamic viscosity

$\nu$     =   kinematic viscosity

$\tau$     =   shear stress

## Introduction

THE Blasius solution of the laminar boundary layer over a thin flat plate is one of the classical exact solution of the Navier-Stokes equations, its result had been exhaustively proven experimentally for more than one hundred years and it represents as well often the first testing case to validate the Computational Fluid Dynamics Software's, yet it shows more than limitation in some range of application [1]. One of these limitations is its inability to detect the leading edge flow as it represents a singularity to the boundary layer equations.

A review of literature reveals that rare studies were addressing this subject as the Blasius solution was satisfactory in obtaining the experimental velocity profile downstream the leading edge.

The current study is analyzing the leading edge singularity by studying the kinematic features of the leading edge average velocity profile, it was found that the leading edge velocity profile is in fact a typical Gaussian distribution curve that grows both in time and space.

## I. THE GAUSSIAN PATTERN: A REVIEW OF THE LITERATURE

The Gaussian profile did appear in several flow situations in the laminar regime as well as in turbulence; as a vast subject the following brief review is meant to illustrate some known cases by way of example.

Oseen & Lamb [2-3] deduced an expression to describe a viscous vortex flow with a maximum vorticity at the center of the vortex, while asymptotically decaying radially to the far field periphery; their approach was based on the diffusion equation where the vorticity is the diffused variable; ending up with an unsteady Gaussian formula describing the vorticity behavior in time and space.

In 1933 Goldstein[4] calculated numerically the velocity profile at the wake behind a flat plate placed parallel to the stream flow, at some considerable distance downstream the velocity profile  was found to have a Gaussian pattern; a prediction that was proven experimentally by Fage and Falkner[4], and later on by Sato and Kuriki[5].

Recently Weyburne [6] observed that the second derivative of the velocity profile and thermal profile in the laminar flat plate boundary layer has a Gaussian-like behavior, and developed a new expression of the boundary layer thickness that is not based on the arbitrary 99% of the stream velocity traditional limit, but instead using the standard probability distribution function methodology.

In the jet turbulence flow, the Gaussian distribution is a widespread pattern; for example: the time-averaged velocity profile in a round jet, in the plane (or slot) jet and in the round plume flow are self-similar (affine) velocity profiles that can be well approximated by the Gaussian curve. On the round buoyant plumes flow Lee and Chu [7] pointed out: "Experiments have shown that the plume is a boundary-layer type of flow. The velocity and the concentration profiles in the fully established flow are similar in shape at all heights, and well-described by Gaussian profiles"; a statement that supports clearly the relation between the boundary layer flow and the Gaussian pattern based on a physical or experimental observation.

## II. THE LAGRANGIAN PERSPECTIVE

The Lagrangian approach in fluid mechanics is characterized by tracking the fluid parcels in time and space, these fluid parcels are specified and identified all along its path through the flow; unlike the conventional Eulerian approach, based on its field perspective, describes the flow at certain specific stations regardless of the identity of parcels that pass that location.

In the current study the Lagrangian approach is adopted where the identified fluid parcels located at the plate leading edge Timelines are monitored in space and time as the flat plate moves through the stationary fluid; the Timelines illustrated by the hydrogen bubble technique in (fig. 1) [8-9] are in fact telling us the history by showing snap shots of the identified parcels during the flow; at first the velocity profile is highly compressed near the wall but as time progresses the diffusion increases and the profile scale gradually grows normal to the plate.

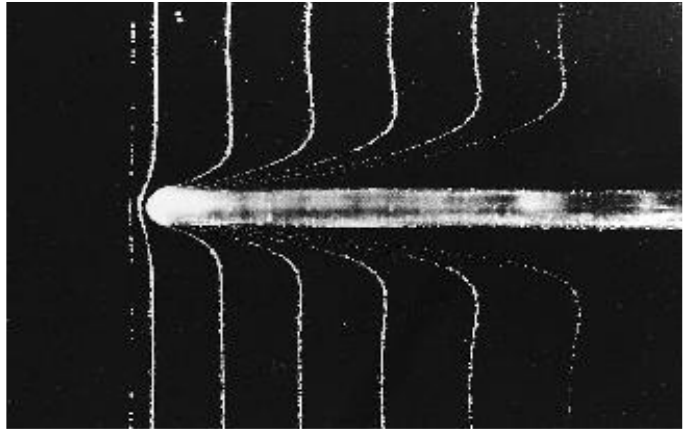

**FIG 1.   Timelines of the leading edge average velocity profile [8]**
**Reproduced with permission from Y. Nakayama and R.F. Boucher, Introduction to Fluid Mechanics**
**(Elsevier, 1998). Copyright 1998, Elsevier.**

### III. THE MOVING PLATE APPROACH: THE THEORETICAL ANALYSIS

It is known obviously that a moving flat plate in a stationary fluid is exactly the same case as a fluid moving over a stationary flat plate; by considering the moving flat plate perspective and adopting a parallel flow approach, i.e. the transversal velocity equals zero at the wall to the infinity far field, our case could be described by means of the one-dimensional diffusion equation eq. (1) where the quantity to be diffused is the flat plate velocity $U$ (fig. 2): at first, only the fluid near the moving plate will be drawn into motion, but as time progresses the thickness of this moving region will increase[10].

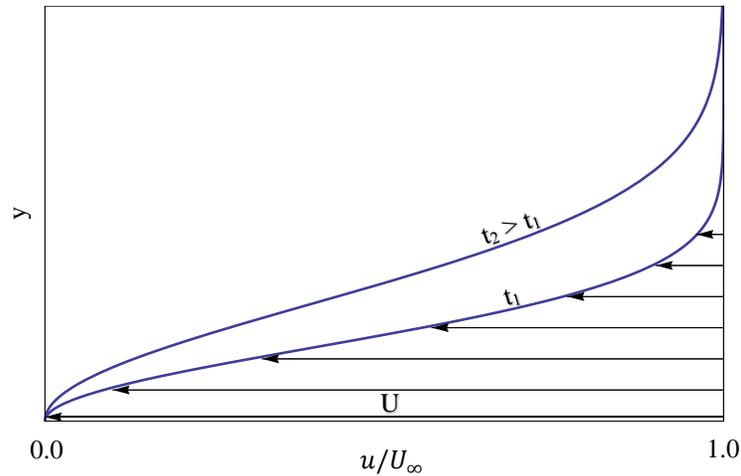

**FIG 2.Diffusion of the plate velocity $U$ & the growth of the leading edge velocity profile in time**

It has to be noted that the diffusion equation was also adopted at the Stokes' First problem that studies the sudden moved flat plate in a stationary fluid [10-11].

By applying the boundary conditions: u(t, $y = \infty$ ) = 0 and u(t > 0, y = 0 ) = U to the diffusion equation eq. (1)

$$\frac{\partial u}{\partial t} = \nu \frac{\partial^2 u}{\partial y^2} \tag{1}$$

and considering the Buckingham Pi theory eq. (2) where $u$ at the first dimensionless term eq. (3) doesn't refer to $U_\infty$, as the Stokes' First problem, but rather is referring to the term $\sqrt{t}/\sqrt{\nu}$.

$$\Pi_1 = f(\Pi_2) \tag{2}$$

$$\Pi_1 = \frac{u\sqrt{t}}{\sqrt{\nu}} \tag{3}$$

The second dimensionless term $\Pi_2$ or η is

$$\Pi_2 = \frac{y}{\sqrt{\nu t}} \tag{4}$$

$$\eta = \frac{y}{\sqrt{\nu t}} \tag{5}$$

By expanding eq. (2) the velocity equation is then

$$u = \frac{\sqrt{\nu}}{\sqrt{t}} f(\eta) \tag{6}$$

The left hand side term of eq. (1) will be: (See Appendix A1 for the development.)

$$\frac{\partial u}{\partial t} = -\frac{1}{2} \frac{\sqrt{\nu}}{t^{1.5}} \left( f(\eta) + \eta \frac{\partial f(\eta)}{\partial \eta} \right) \tag{7}$$

The right hand side term of eq. (1) will be: (See Appendix A2 for the development.)

$$\nu \frac{\partial^2 u}{\partial y^2} = \nu \frac{1}{\sqrt{\nu} t^{1.5}} \frac{\partial^2 f(\eta)}{\partial \eta^2} \tag{8}$$

Then eq. (1) reads

$$-\frac{1}{2} \frac{\sqrt{\nu}}{t^{1.5}} \left( \partial f(\eta) + \eta \frac{\partial f(\eta)}{\partial \eta} \right) = \nu \frac{1}{\sqrt{\nu} t^{1.5}} \frac{\partial^2 f(\eta)}{\partial \eta^2} \tag{9}$$

To obtain the ordinary differential

$$\frac{d^2 f(\eta)}{d\eta^2} + \frac{1}{2}\left(\frac{d\eta\, f(\eta)}{d\eta}\right) = 0 \tag{10}$$

Arranging and integrating once

$$\frac{df(\eta)}{d\eta} + \frac{\eta\, f(\eta)}{2} = C_0 \tag{11}$$

To find $C_0$ the boundary condition at the far field is considered: at $y = \infty$ ; u = 0 therefore f($\eta$) = 0 as per eq. (6),

with $C_0 = 0$ we have a homogeneous ordinary differential equation whose solution is

$$f(\eta) = C_1\, e^{\frac{-\eta^2}{4}} \tag{12}$$

Substituting eq. (12) into eq. (6) we obtain

$$u = \frac{\sqrt{\nu}}{\sqrt{t}} C_1\, e^{-\frac{y^2}{4\nu t}} \tag{13}$$

Putting  y = 0 to obtain the flat plate velocity U (or $U_\infty$ in another perspective)

$$U = \frac{\sqrt{\nu}}{\sqrt{t}} C_1 \tag{14}$$

To end up with the formula of a Gaussian distribution velocity profile

$$\frac{u}{U} = e^{-\frac{y^2}{4\nu t}} \tag{15}$$

## IV. PLOTTING THE TIMELINES

The normalized average velocity profile at the leading edge is then:

$$\frac{u}{U_\infty} = e^{-\frac{y^2}{4\nu t}} \tag{16a}$$

$$= e^{-\frac{y^2 U_\infty}{4\nu x}} \tag{16b}$$

The plotting of several leading edge average velocity profiles, based on eq. (16) are shown at (fig. 3). The profiles are affine or similar to one another and growing by factor of $\sqrt{x}$ in the perpendicular direction to the flat plate which is a main feature of the flat plate laminar boundary layer: for example the dashed curve is the locus of the boundary layer thickness, or $\delta_{0.99}$, where $\frac{u}{U_\infty} = 0.99$.

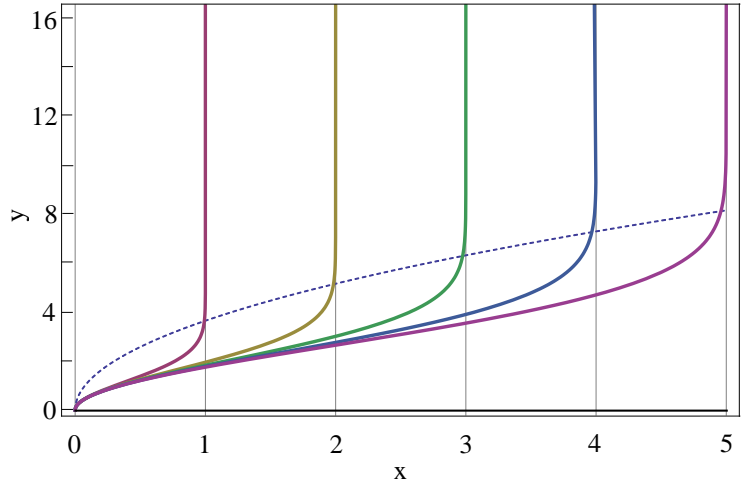

**FIG 3. Simulation of Time Lines of the leading edge average velocity profiles**

## V. THE LEADING EDGE VELOCITY PROFILE VERSUS THE BLASIUS PROFILE

The Blasius values (coordinates) [11] mapping the laminar velocity profile are plotted versus the Gaussian curve (fig. 4) confirm a perfect match especially over the 90% of the top part of the curve; *thus demonstrating that the Blasius profile is in fact a truncated curve of a Gaussian nature.* The reason the bottom part of the curve (near-to-wall) is difficult to visualize during experiments is that the velocity profile is *highly compressed* at low Reynolds number (with respect to x); going downstream, as the Reynolds number increases the boundary layer thickness increases consequently stretching the velocity profile at the normal direction, therefore it is easier to practically visualize the bottom part of the Gaussian profile at higher Reynolds number than at lower Reynolds number.

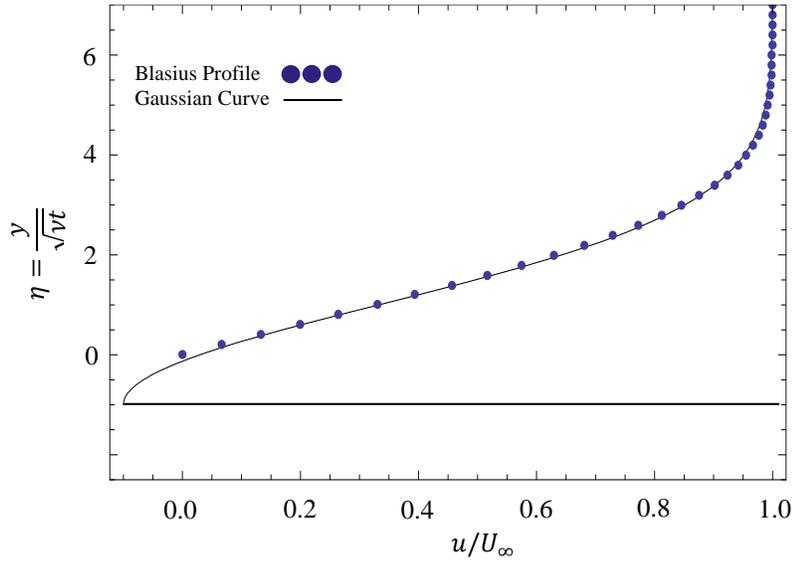

**FIG 4.** **Blasius Profile (Points) versus the Gaussian curve**

Measuring the bottom part (near-to-the wall part) of the profile is discussed in an experiment [12] (fig.5) where Klewicki et al. pointed out: *"the readings are higher than the expected values of the Blasius profile, and to compensate for this effect, stop measurements at* $u/U_\infty \approx 0.1$, *and for mean flow, use linear extrapolation to the wall"* [12].

Therefore, in order to solve this paradox they attribute the deviation from the theory to a certain external influence; they explain: "*As the hot wire gets closer to the wall, radiation from the model removes heat from the hot wire, resulting in readings of higher velocity than is actually present.*"

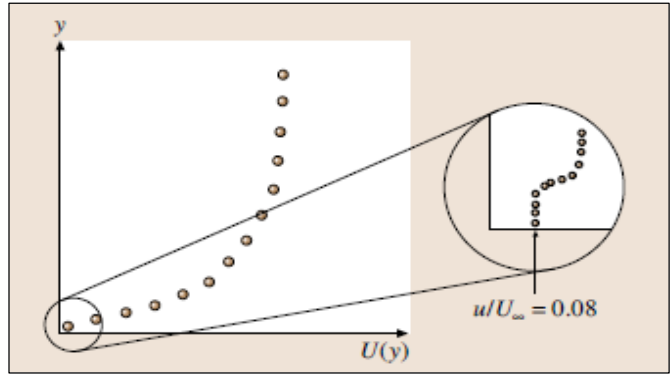

**FIG 5. Measuring the "near to the wall" part of the downstream velocity profile [12]**
**Reproduced with permission from Springer, ©Springer-Verlag Berlin Heidelberg 2007.**

In fact and according to the current study the so called higher than expected values that were measured experimentally, shown at the right zoomed view (fig. 5), are *the true and correct values of the Gaussian curve* which has a zero velocity gradient at the wall.

## VI. RELATION BETWEEN THE LEADING EDGE VELOCITY PROFILE AND THE DOWNSTREAM VELOCITY PROFILE

Knowing the formula of the leading edge velocity profiles the downstream local velocity profiles, in a parallel flow, can be obtained by *subtracting two successive leading edge profiles spaced by an interval of time: $\Delta t$* (fig.6), for instance the solid velocity profile (c) at the right is the result of subtraction of the two dashed leading edge velocity profiles (a) and (b), and similarly for the other two solid profiles; these are actually the velocity profiles that are measured and partly visualized during experiments.

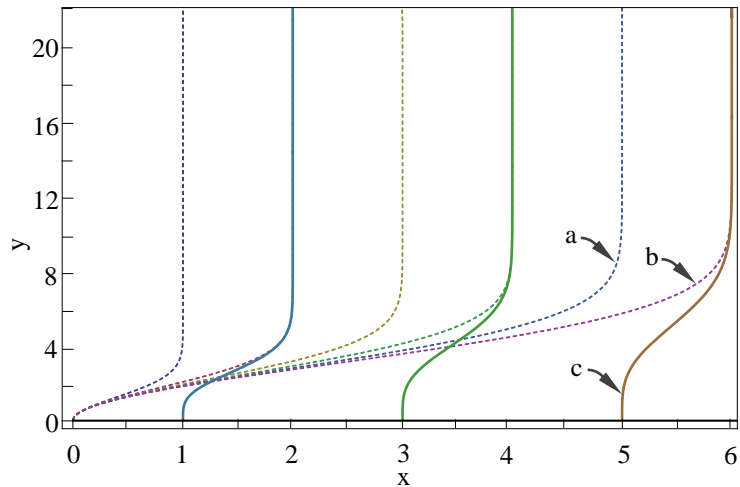

**FIG 6. The leading edge velocity profiles (dashed) and the downstream velocity profiles (solid)**

It has to be emphasized that the leading edge velocity profiles are not limited to low Reynolds number (with respect to x); thus the definition of the leading edge velocity profile is: *it is the velocity profile that has one extremity stagnating at the flat plate leading edge whereas the other extremity is flowing freely with the Timelines free stream velocity*. Therefore the local velocity profile at a certain x is:

$$\frac{u}{U_\infty} = x_{(t+\Delta t)} \, e^{-\frac{y^2}{4\nu(t+\Delta t)}} \;-\; x_{(t)} \, e^{-\frac{y^2}{4\nu t}} \tag{17}$$

The downstream velocity profiles are following the same trend as its parents i.e. the leading edge profiles, in terms of the perpendicular growth factor $\sqrt{x}$, whereas the shape of the profiles are more flattened near the wall comparing to the leading edge profiles. In fact the downstream profiles represent the Eulerian description of the flow as it reflect the particles velocity distribution at a specific location, whereas the leading edge profiles represent the Lagrangian description of the flow as they are tracking the path lines of *specific particles* moving from the start of time [13-14].

## VII. THE DISPLACEMENT THICKNESS

Shlichting, in his landmark book [11], considered that the exact calculation of the displacement thickness as mission impossible due to the fact that the boundary layer has an ambiguous limit; it is chosen arbitrary to be 99% of the free stream velocity (or 95% [9]), yet based on the exact formula of the leading edge velocity profile, finding the displacement thickness at a certain distance from the leading edge is a straightforward calculation.

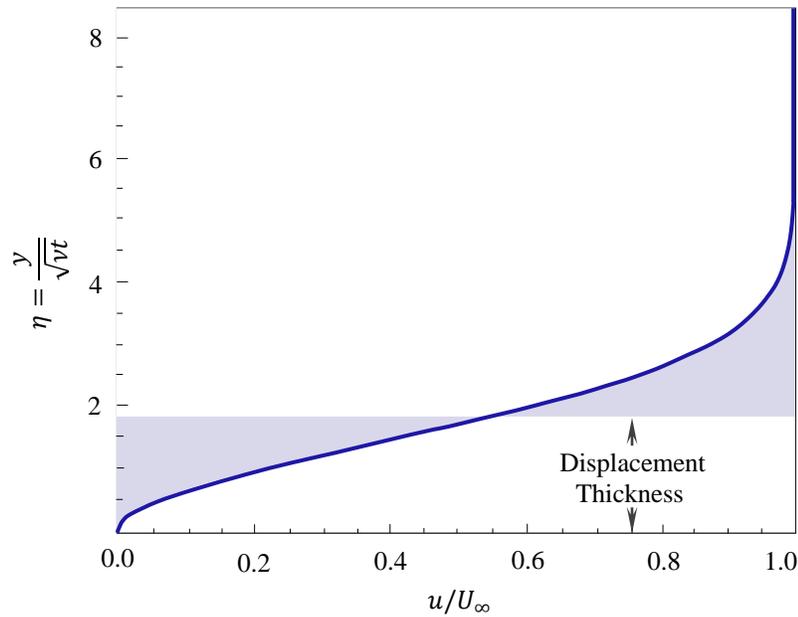

**FIG 7. The normalized leading edge velocity profile and its displacement thickness**

The Gaussian distribution curve has a finite integral to calculate the whole area under the curve: "The Error Function", whereas there is no integral formula to calculate a *part* of that area, therefore the known finite integral should have the limits 0 to ∞ or −∞ to ∞.

Thus the area under the normalized curve (fig. 7) i.e. the displacement thickness is:

$$\delta = \int_0^\infty e^{-\frac{y^2}{4\nu t}} \; dy \tag{18a}$$

$$= \sqrt{\pi \nu t} \tag{18b}$$

$$\delta = 1.772\sqrt{\nu}\sqrt{\frac{x}{U_\infty}} \tag{19}$$

The constant found by Blasius solution for computing the displacement thickness which is 1.7208 [11] is now understood its origin as $\sqrt{\pi}$; it is lower slightly comparing to our value since Blasius solution approximates the area calculation by stopping at $u/U_\infty = 0.99$ whereas our calculation, based on the exact formula of the leading edge velocity profile, takes the whole area into account.

## VIII. THE DRAG FORCE ANALYSIS

In his solution Blasius calculated the drag by finding the wall shear stress of the flat plate surface Eq. (20), yet according to the current study the velocity gradient $\frac{\partial u}{\partial y}$ at the flat plate level is zero, therefore *the conventional shear stress formula cannot be applicable as it gives a trivial null shear stress*; another geometrical property of the velocity profile has to be used in order to calculate the resisting shear force.

$$\tau = \mu \frac{\partial u}{\partial y} \tag{20}$$

The displacement thickness is an instantaneous parameter that describes how far the flow is drifted away normal to the flat plate at a certain time, in other terms it describes how far the leading edge velocity profile is stretched away normal to the flat plate, meanwhile as the flow advances downstream the leading edge velocity profile is naturally subjected to another expansion parallel to the flat plate, therefore the perpendicular displacement thickness combined with the profile parallel displacement constitute a *displacement area*; that parameter is representing the loss that the flow is subjected to, or in other term it is a function of the actual shear stress that the plate *feels* during its movement through the flow.

As the flat plate starts moving from rest and by adopting a linear relation between the flat plate velocity and time, the *average* velocity $U$ is then equal to the actual flat plate velocity at $t/2$, where $x_{(t/2)} = 1/4\ x_{(t)}$.

Then the total displacement, or the *displacement area* in other term is:

$$\delta_T = \sqrt{\pi}\ \sqrt{\nu}\ \sqrt{\frac{x}{U_\infty}} \cdot \frac{1}{4}\ x \tag{21}$$

In terms of time:

$$\delta_T = \sqrt{\pi}\ \sqrt{\nu}\ \sqrt{t}\ \ \frac{1}{4}\ U_\infty t \tag{22}$$

Differentiating with respect to time to obtain the rate of the total displacement growth:

$$\frac{1.5}{4}\sqrt{\pi} U_\infty \sqrt{\nu}\sqrt{t}\ = 0.664\ U_\infty \sqrt{\nu}\sqrt{t} \tag{23}$$

Differentiating a second time to obtain the instantaneous acceleration of the total displacement growth:

$$a_{inst.} = 0.332\ U_\infty \sqrt{\nu}\frac{1}{\sqrt{t}} \tag{24}$$

In terms of x:

$$a_{inst.} = 0.332\ U_\infty \sqrt{\nu}\sqrt{\frac{U_\infty}{x}} \tag{25}$$

The total acceleration over a length x is:

$$a = \int_0^x 0.332\ U_\infty \sqrt{\nu}\sqrt{\frac{U_\infty}{x}}\ dx \tag{26}$$

$$a = 0.664\ \sqrt{\nu}\ \sqrt{x}\ U_\infty{}^{1.5} \tag{27}$$

Then the total drag force over a length x on one side of the flat plate is:

$$D = 0.664 \sqrt{\rho} \sqrt{\mu} \sqrt{x} \, U_{\infty}^{1.5} \tag{28}$$

Which is the same formula of the drag force on one side of the plate that Blasius obtained.

Similarly as shown previously for the displacement thickness constant, the constant found by Blasius for computing the drag force that was proven experimentally which is 0.664 [11] is now understood its origin as $\frac{1.5}{4}\sqrt{\pi}$.

## IX. CONCLUSION

A new Lagrangian analysis of the laminar thin flat plate boundary layer is performed based on the diffusion equation reveals that the leading edge velocity profiles are similar or affine Gaussian distribution curves that grows in time and space. The Blasius velocity profile represents approximately 90% of the Gaussian curve. The kinematic observation as well as the theoretical study prove that the Gaussian curve is the real shape of the leading edge profile; an experimental study employing an *ultra-thin* flat plate is to be conducted in order to obtain the final confirmation.

As for the shear stress calculation; a new approach to calculate the drag force is adopted based on the displacement area concept of the growing leading edge velocity profile rather than the gradient of the downstream velocity profiles at the wall.

On redefining the Steady Flow; the leading edge profiles are unsteady profiles by nature as its formula illustrates, despite the fact that the free stream velocity is apparently steady. The parabolic growth of the downstream velocity profiles as $x$ increases is in fact due to the unsteadiness nature of the leading edge velocity profiles at the transversal direction.

The downstream velocity profiles at a certain $x$, based on the parallel flow approach, can be obtained by subtracting two successive leading edge velocity profiles.

## X. APPENDIX A1: Development of $\frac{\partial u}{\partial t}$ :

$$\frac{\partial u}{\partial t} = \frac{\partial \left( \frac{\sqrt{\nu}}{\sqrt{t}} \, f(\eta) \right)}{\partial t} \qquad ; \qquad \frac{\partial u}{\partial t} = -\frac{1}{2} \frac{\sqrt{\nu}}{t^{1.5}} \, f(\eta) + \frac{\sqrt{\nu}}{\sqrt{t}} \, \frac{\partial f(\eta)}{\partial \eta} \frac{\partial \eta}{\partial t}$$

$$\frac{\partial \eta}{\partial t} = -\frac{1}{2} \frac{y}{\sqrt{\nu} \, t^{1.5}} = \frac{-\eta}{2t} \quad ; \quad \frac{\partial u}{\partial t} = -\frac{1}{2} \frac{\sqrt{\nu}}{t^{1.5}} \, f(\eta) - \frac{\sqrt{\nu}}{\sqrt{t}} \, \frac{\partial f(\eta)}{\partial \eta} \frac{\eta}{2t}$$

$$\frac{\partial u}{\partial t} = -\frac{1}{2} \frac{\sqrt{\nu}}{t^{1.5}} \, \left( f(\eta) + \eta \frac{\partial f(\eta)}{\partial \eta} \right)$$

## X. APPENDIX A2: Development of $\frac{\partial^2 u}{\partial y^2}$ :

$$\frac{\partial u}{\partial y} = \frac{\partial \left( \frac{\sqrt{\nu}}{\sqrt{t}} \, f(\eta) \right)}{\partial y} \; ; \; \frac{\partial u}{\partial y} = \frac{\sqrt{\nu}}{\sqrt{t}} \left( \frac{\partial f(\eta)}{\partial \eta} \frac{\partial \eta}{\partial y} \right) \; ; \; \frac{\partial u}{\partial y} = \frac{\sqrt{\nu}}{\sqrt{t}} \left( \frac{\partial f(\eta)}{\partial \eta} \frac{1}{\sqrt{\nu t}} \right) \; ; \; \frac{\partial u}{\partial y} = \frac{1}{t} \left( \frac{\partial f(\eta)}{\partial \eta} \right)$$

$$\frac{\partial^2 u}{\partial y^2} = \frac{\partial \left( \frac{1}{t} \frac{\partial f(\eta)}{\partial \eta} \right)}{\partial y} \; ; \; \frac{\partial^2 u}{\partial y^2} = \frac{\partial \left( \frac{1}{t} \frac{\partial f(\eta)}{\partial \eta} \right)}{\partial \eta} \frac{\partial \eta}{\partial y} \; ; \; \frac{\partial^2 u}{\partial y^2} = \frac{1}{t} \frac{\partial^2 f(\eta)}{\partial \eta^2} \frac{1}{\sqrt{\nu t}}$$